\documentclass[12pt]{iopart}

\usepackage{cite}

\begin{document}

\title[Antiunitary symmetry]{Generalization of parity-time and partial parity-time symmetry}
\author{Francisco M. Fern\'{a}ndez}
\address{INIFTA (UNLP, CCT La Plata-CONICET), Blvd. 113 y 64 S/N,
Sucursal 4, Casilla de Correo 16, 1900 La Plata,
Argentina}\ead{fernande@quimica.unlp.edu.ar}

\maketitle

\begin{abstract}
We show that parity-time and partial parity-time symmetries are particular
cases of antiunitary symmetry. This point is illustrated by means of a
recently discussed system of non-Hermitian coupled harmonic oscillators that
also exhibits other types of antiunitary symmetries. We also show that a
combination of group and perturbation theory is a useful tool for predicting
broken antiunitary symmetry.
\end{abstract}

\section{Introduction}

\label{sec:intro}

In a recent series of papers Fern\'{a}ndez and Garcia\cite{FG14b,FG14c} and
Amore et al\cite{AFG14b,AFG15} applied group theory to non-Hermitian coupled
oscillators with the purpose of determining the conditions under which the
well known parity-time (PT) symmetry\cite{B07} as well as its generalization
space-time (ST) symmetry\cite{KC08} are unbroken. One of the examples
discussed by Fern\'{a}ndez and Garcia\cite{FG14b} is a pair of harmonic
oscillators coupled by an imaginary quadratic interaction $%
H_{2D}=p_{x}^{2}+p_{y}^{2}+\omega _{x}^{2}x^{2}+\omega _{y}^{2}y^{2}+iaxy$.
They found that when the two oscillators are identical ($\omega
_{x}^{2}=\omega _{y}^{2}$) the eigenvalues $E_{mn}$ are real when $m=n$ and
pairs of complex conjugate numbers when $m\neq n$. On the other hand, when $%
\omega _{x}^{2}\neq \omega _{y}^{2}$ all the eigenvalues are real for
certain combinations of the oscillator parameters $\omega _{x}^{2}$, $\omega
_{y}^{2}$ and $a$. That is to say: in the latter case there is a sort of
phase transition from real to complex eigenvalues.

More recently, Beygi et al\cite{BKB15} discussed a family of $N$ coupled
oscillators that contain the example discussed by Fern\'{a}ndez and Garcia
when $N=2$. These authors assumed that each oscillator interacts only with
its nearest neighbors in a sort of linear chain; that is to say: the $i$-th
oscillator interacts only with the $(i-1)$-th and $(i+1)$-th ones. They
studied both quantum-mechanical as well as classical oscillators and arrived
at exactly the same conclusions drawn earlier by Fern\'{a}ndez and Garcia%
\cite{FG14b} for the former case. According to Beygi et al those
non-Hermitian coupled harmonic oscillators exhibit partial PT
symmetry which is a concept introduced somewhat earlier by
Yang\cite{Y14}. Both PT symmetry and partial PT symmetry are
particular cases of the ST symmetry introduced by Klainman and
Cederbaun\cite{KC08} and all of them are examples of the
antiunitary symmetry discussed by Wigner\cite{W60} long time
before. It is well known that if $U$ is a unitary transformation
in configuration space and $T$ is the time reversal operation then
$A_{U}=UT$ is an antiunitary operator\cite{W60}. Therefore, it
seems more reasonable to speak of antiunitary symmetry as the most
general concept. When $A_{U}$ leaves the non-Hermitian Hamiltonian
operator $H$ invariant ($A_{U}HA_{U}^{-1}=H$) we say that $H$
exhibits antiunitary symmetry. In the particular case that $U$ is
the parity operation $P$ then we are in the presence of PT
symmetry. According to Yang\cite{Y14} when the potential energy
function $V(x,y)$ of a
two-dimensional quantum-mechanical model satisfies $V(-x,y)^{*}=V(x,y)$ or $%
V(x,-y)^{*}=V(x,y)$ then the model satisfies partial PT symmetry (note that $%
V(-x,-y)^{*}=V(x,y)$ describes the usual PT symmetry). Obviously, it is a
particular case of the antiunitary symmetry generated by the unitary
operators $U_{x}:(x,y)\rightarrow (-x,y)$ or $U_{y}:(x,y)\rightarrow (x,-y)$%
, respectively. Note, for example, that the Hamiltonian $H_{2D}$ is
invariant with respect to the antiunitary operators $A_{x}=U_{x}T$ and $%
A_{y}=U_{y}T$. In fact, Klainman and Cederbaum\cite{KC08} chose $%
H=H_{0}+i\lambda xy$ as a simple prototypical example of ST symmetry. The
exactly-solvable $N$-dimensional harmonic oscillators chosen by Beygi et al%
\cite{BKB15} that exhibit partial PT symmetry are also particular examples
of antiunitary symmetry.

By means of group theory and perturbation theory Amore et al\cite{AFG14b}
conjectured that of all the examples of antiunitary symmetry studied so far,
PT symmetry appears to be the less likely to be broken. Their conclusion was
based on the role of $U$ in the symmetry group $G_{0}$ for the unperturbed
Hermitian oscillator $H_{0}$.

In a later communication Fern\'{a}ndez and Garcia\cite{FG15} discussed three
PT-symmetric models with completely different spectra. They are of the form $%
H=H_{0}+igz$, where $H_{0}$ is an Hermitian operator with a central-field
potential $V(r)$. The first one is exactly solvable and exhibits a real
spectrum for all values of $g$. On the other side, the second one exhibits
complex eigenvalues for all values of $g$. Finally, with an intermediate
behavior, the third one shows the well known phase transition typical of
PT-symmetric quantum-mechanical models\cite{BW12}. The second model is most
interesting because the search carried out by Fern\'{a}ndez and Garcia\cite
{FG14b,FG14c} and Amore et al\cite{AFG14b,AFG15} failed to produce any
PT-symmetric Hamiltonian with completely broken PT symmetry. In this case
the phase transition takes place at the trivial Hermitian limit $g=0$. The
remarkable difference in the spectra of those non-Hermitian operators can be
traced back to the symmetry of $H_{0}$. The higher this symmetry the more
probable the appearance of complex eigenvalues\cite{FG15}.

One of the aims of this paper is to analyse the system of coupled
oscillators proposed by Beygi et al\cite{BKB15} from the point of view of
group theory and perturbation theory. In Sec.~\ref{sec:A_symmetry} we
develop the concept of antiunitary symmetry in a way that generalizes and
extends the original idea of Klainman and Cederbaum\cite{KC08} and improves
the point-group analysis carried out by Amore et al\cite{AFG14b}. We also
show that the combination of group theory and perturbation theory yields a
fairly good idea about the kind of spectrum that one expects of a given
non-Hermitian operator. In Sec.~\ref{sec:harm_osc} we study the $N$%
-dimensional coupled harmonic oscillators of Beygi et al\cite
{BKB15} from the point of view of point-group theory. In addition
to finding the point group for these oscillators we show that
their antiunitary symmetry is greater than the partial PT symmetry
discussed so far. Finally, in Sec.~\ref{sec:conclusions} we
summarize the main results of the paper and draw conclusions.

\section{Antiunitary symmetry}

\label{sec:A_symmetry}

Consider a Hamiltonian operator of the form
\begin{equation}
H(\lambda )=H_{0}+\lambda H^{\prime },  \label{eq:H_general}
\end{equation}
where $H_{0}$ is Hermitian and $H^{\prime }$ is real and linear. Suppose
that there is a group $G=\left\{ U_{1},U_{2},\ldots ,U_{n}\right\} $ of
unitary operators ($U_{i}^{\dagger }=U_{i}^{-1}$) that leave both $H_{0}$
and $H^{\prime }$ invariant:
\begin{equation}
U_{i}H_{0}U_{i}^{-1}=H_{0},\;U_{i}H^{\prime }U_{i}^{-1}=H^{\prime }.
\label{eq:G_for_H}
\end{equation}
This group of operators is commonly called the symmetry group for $H$\cite
{C90,T64}. Suppose that there is a set of unitary operators $S_{W}=$ $%
\left\{ W_{1},W_{2},\ldots ,W_{m}\right\} $ such that
\begin{equation}
W_{i}H_{0}W_{i}^{-1}=H_{0},\;W_{i}H^{\prime }W_{i}^{-1}=-H^{\prime }.
\label{eq:S_W}
\end{equation}
Since $U_{i}W_{j}\in S_{W}$ and $W_{i}W_{j}\in G$ we conclude that $%
G_{0}=G\cup S_{W}$ is a group of unitary operators that leave $H_{0}$
invariant. Obviously, $G_{0}$ is at least a subgroup of the symmetry group
for $H_{0}$.

Let us now consider the set of antiunitary operators $S_{A}=$ $\left\{
A_{1},A_{2},\ldots ,A_{m}\right\} $, where $A_{i}=W_{i}T=TW_{i}$ and $T$ is
the time-reversal operator\cite{W60}. If $\lambda ^{*}=-\lambda $ then $%
A_{i}H(\lambda )A_{i}^{-1}=H(-\lambda ^{*})=H(\lambda )$. Besides, since $%
U_{i}A_{j}\in S_{A}$ and $A_{i}A_{j}\in G$ we conclude that $G_{A}=G\cup
S_{A}$ is a group of operators that leave $H(\lambda )$ invariant
\begin{equation}
KH(\lambda )K^{-1}=H(\lambda ),\;K\in G_{A},\;\lambda ^{*}=-\lambda .
\label{eq:G_ST}
\end{equation}
We call $G_{A}$ the antiunitary-symmetry group for $H$ as a generalization
of the concept introduced by Klainman and Cederbaum\cite{KC08}. In this
sense, PT symmetry is a particular case of antiunitary symmetry where $P\in
S_{W}$, $P:\mathbf{x}\rightarrow -\mathbf{x}$. Another particular case of
antiunitary symmetry is the partial PT symmetry introduced by Yang\cite{Y14}
which for two-dimensional models takes the form $W_{x}:(x,y)\rightarrow
(-x,y)$ or $W_{y}:(x,y)\rightarrow (x,-y)$. As indicated in the introduction
Klainman and Cederbaum\cite{KC08} already chose these coordinate
transformations to illustrate the concept of ST symmetry (they called them $%
P_{x}$ and $P_{y}$, respectively). This kind of symmetry was recently
extended by Beygi et al\cite{BKB15} to $N$-dimensional oscillators.

If we apply $W\in S_{W}$ to the eigenvalue equation
\begin{equation}
H(\lambda )\psi _{n}=E_{n}(\lambda )\psi _{n},  \label{eq:Schr_gen}
\end{equation}
we have
\begin{equation}
WH(\lambda )W^{-1}W\psi _{n}=H(-\lambda )W\psi _{n}=E_{n}(\lambda )W\psi
_{n}.  \label{eq:SHpsi}
\end{equation}
It is clear that
\begin{equation}
E_{n}(\lambda )=E_{m}(-\lambda ),  \label{eq:En(lam)=Em(-lam)}
\end{equation}
where $E_{m}(-\lambda )$ is an eigenvalue of $H(-\lambda )$. Since this
equation should be valid for all $\lambda $, then when $\lambda \rightarrow
0 $ we have $E_{n}(0)=E_{m}(0)$. If the eigenvalue $E_{n}(0)$ of $H_{0}$ is
nondegenerate, then $m=n$, $E_{n}(-\lambda )=E_{n}(\lambda )$ and
perturbation theory yields the formal power series
\begin{equation}
E_{n}(\lambda )=\sum_{j=0}^{\infty }E_{n}^{(j)}\lambda ^{2j}.
\label{eq:En_series_even}
\end{equation}
Klaiman et al\cite{KGM08} obtained a similar result by means of a lengthier
argument. If the radius of convergence of this series is finite, then we
conclude that $E_{n}(\lambda )$ is real for sufficiently small $|\lambda |$
when $\lambda =-\lambda ^{*}$. If, on the other hand, the eigenvalue $%
E_{n}(0)$ is degenerate, then the perturbation series for
$E_{n}(\lambda )$ may exhibit odd powers of $\lambda $ and one
expects complex eigenvalues under such conditions. This argument
is also expected to apply to the more general case of a divergent
perturbation series provided it is asymptotic to $E_{n}(\lambda )$
because we only have to consider sufficiently small values of
$|\lambda |$ to prove that the eigenvalue is complex. We
appreciate that the occurrence of real eigenvalues of a
non-Hermitian operator $H(ig)$, $g$ real, is strongly dependent on
the form of the spectrum of $H_{0}$. This approach based on
perturbation theory is not new\cite{AFG14b} but we outline it here
for completeness.

\section{Coupled harmonic oscillators}

\label{sec:harm_osc}

In this section we focus our attention on the set of adjacently coupled
harmonic oscillators studied by Beygi et al\cite{BKB15}
\begin{equation}
H(\lambda )=\frac{1}{2}\sum_{i=1}^{N}\left( p_{i}^{2}+\omega
_{i}^{2}x_{i}^{2}\right) +\lambda \sum_{j=1}^{N-1}x_{j}x_{j+1},
\label{eq:H_CHO}
\end{equation}
where $p_{j}=-i\partial /\partial x_{j}$ in the coordinate representation.
As indicated in the introduction, the particular case $N=2$ was studied
earlier by Fern\'{a}ndez and Garcia\cite{FG14b}. If we choose the oscillator
frequencies $\omega _{j}$ so that the spectrum of $H_{0}=H(0)$ is
nondegenerate then, arguing as in Sec.~\ref{sec:A_symmetry}, the spectrum of
$H(ig)$ is expected to be real for sufficiently small values of $g$. When
some of the frequencies are equal, degeneracy emerges as well as the chance
of complex eigenvalues. Consequently, one expects phase transitions; that is
to say surfaces in the $\omega $-plane that separate regions of real and
complex eigenvalues. Fern\'{a}ndez and Garcia\cite{FG14b} and more generally
and exhaustively Beygi et al\cite{BKB15} discussed some particular cases in
detail.

From the point of view of symmetry the case of equal frequencies is the most
interesting one; therefore in what follows we choose $\omega _{j}=1$ for all
$j$. We first obtain the unitary operators $U_{j}$ that leave $H^{\prime }$
invariant. They are given by the coordinate transformations
\begin{eqnarray}
U_{1} &:&(x_{1},x_{2},\ldots ,x_{N})\rightarrow (x_{1},x_{2},\ldots ,x_{N}),
\nonumber \\
U_{2} &:&(x_{1},x_{2},\ldots ,x_{N})\rightarrow (-x_{1},-x_{2},\ldots
,-x_{N}),  \nonumber \\
U_{3} &:&(x_{1},x_{2},\ldots ,x_{N})\rightarrow (x_{N},x_{N-1},\ldots
,x_{1}),  \nonumber \\
U_{4} &:&(x_{1},x_{2},\ldots ,x_{N})\rightarrow (-x_{N},-x_{N-1},\ldots
,-x_{1}).  \label{eq:Uj_group}
\end{eqnarray}
The set $G_{4}=\left\{ U_{1},U_{2},U_{3},U_{4}\right\} $ is an
Abelian group isomorphic to $D_{2}$, $C_{2v}$ and
$C_{2h}$\cite{C90,T64}.

In addition to these four operators there are other four ones that change
the sign of $H^{\prime }$ leaving $H_{0}$ invariant:
\begin{eqnarray}
W_{1} &:&x_{j}\rightarrow (-1)^{j}x_{j},  \nonumber \\
W_{2} &:&x_{j}\rightarrow (-1)^{j+1}x_{j},  \nonumber \\
W_{3} &=&U_{3}W_{1},  \nonumber \\
W_{4} &=&U_{3}W_{2}.  \label{eq:S_W_set}
\end{eqnarray}
We can thus construct four antiunitary operators $A_{j}=W_{j}T$, $j=1,2,3,4$
that leave $H(ig)$ invariant and describe its antiunitary symmetry. Beygi et
al\cite{BKB15} only considered $A_{1}$ and $A_{2}$ in their discussion of
partial PT symmetry. If $A_{1}$ and $A_{2}$ describe partial PT symmetry,
how do we call the antiunitary symmetries associated to $A_{3}$ and $A_{4}$:
reverse-order partial PT symmetry? Instead of adding more fancy names to new
antiunitary transformations of the Hamiltonian operator we suggest the
general term antiunitary symmetry.

The set of eight operators $G_{8}=\left\{
U_{1},U_{2},U_{3},U_{4},W_{1},W_{2},W_{3},W_{4}\right\} $ is also
a group, with a structure that depends on $N$. These operators
obviously leave $H_{0} $ invariant because $G_{8}$ is a subgroup
of the actual symmetry group for $H_{0}$ that we do not discuss
here. $G_{8}$ is isomorphic to $C_{4v}$ when $N$ is even and to
$D_{2h}$ when $N$ is odd\cite{C90,T64}.

Each eigenvalue $E_{k}^{(0)}$ of $H_{0}$ is $g_{k}$-fold degenerate, where
\begin{eqnarray}
E_{k}^{(0)} &=&k+\frac{N}{2},\;k=\sum_{j=1}^{N}n_{j},\;n_{j}=0,1,\ldots ,
\nonumber \\
g_{k} &=&\frac{(k+N-1)!}{k!(N-1)!}.  \label{eq:E^0_k,g_k}
\end{eqnarray}
Such great degeneracy is the reason why so many eigenvalues of $H(ig)$ are
complex. For example, when $N=2$ all the eigenvalues with $n_{1}\neq n_{2}$
are complex\cite{FG14b,BKB15}. For any value of $N$ the lowest eigenvalue $%
E_{0}^{(0)}$ is nondegenerate and $E_{0}(ig)$ is real for all $g$\cite{BKB15}
in agreement with the perturbation analysis outlined in Sec.~\ref
{sec:A_symmetry}.

\section{Conclusions}

\label{sec:conclusions}

Instead of inventing new names for new antiunitary symmetries that may
appear in future investigations we suggest to resort to the quite general
concept of antiunitary symmetry. It is perfectly reasonable to keep PT
symmetry for historical reasons but it is far more convenient to avoid
particular names for all its possible variants such as partial PT symmetry,
reverse-order partial PT symmetry, etc. and simply use antiunitary symmetry.
To this end we have tried to formalize the concept of antiunitary symmetry
by means of group theory.

We have also tried to show that perturbation theory is useful for predicting
whether a given non-Hermitian Hamiltonian will exhibit antiunitary symmetry,
at least for some kind of examples. Although the approach is not completely
rigorous it nevertheless provides a fairly good idea of what to expect. As a
general rule, the greater the symmetry of $H_{0}$ the more probable the
occurrence of complex eigenvalues for all values of $g$. This point of view
was used successfully in earlier papers\cite{FG14b,FG14c,AFG14b,AFG15}.

As an illustrative example of the main theoretical ideas we have chosen the
coupled harmonic oscillators studied by Beygi et al\cite{BKB15} and
disclosed their point-group symmetry as well as their antiunitary symmetry.

\end{document}